\documentclass[sort&compress,times,3p,twocolumn,number,12pt]{elsarticle}

\usepackage{graphicx}
\usepackage{times}

\journal{Physics Letters B}
\begin{document}

\begin{frontmatter}
\title{Hard Photodisintegration of a Proton Pair}

% Ph.D. Students
\author[tau]{I.~Pomerantz}
\author[tau]{N.~Bubis}
% Collaboration
\author[ukn]{K.~Allada}
\author[nrcn]{A.~Beck}
\author[nrcn]{S.~Beck}
\author[gwu]{B.~L.~Berman}
\author[fiu]{W.~Boeglin}
\author[tjnaf]{A.~Camsonne}
\author[odu]{M.~Canan}
\author[uva]{K.~Chirapatpimol}
\author[infn]{E.~Cisbani}
\author[infn]{F.~Cusanno}
\author[tjnaf]{C.~W. de ~Jager}
\author[ukn]{C.~Dutta}
\author[infn]{F.~Garibaldi}
\author[uva]{O.~Geagla}
\author[tjnaf,rut]{R.~Gilman}
\author[smu,dal]{J.~Glister}
\author[tjnaf]{D.~W.~Higinbotham}
\author[rut]{X.~Jiang}
\author[ksu]{A.~T.~Katramatou}
\author[ksu]{E.~Khrosinkova}
\author[snu]{B.~W.~Lee}
\author[tjnaf]{J.~J.~LeRose}
\author[uva]{R.~Lindgren}
\author[smu]{E.~McCullough}
\author[tjanf]{D.~Meekins}
\author[tjnaf]{R.~Michaels}
\author[cwm]{B.~Moffit}
\author[ksu]{G.~G. ~Petratos}
\author[tau]{E.~Piasetzky}
\author[duu]{X.~Qian}
\author[mit]{Y.~Qiang}
\author[fiu]{I.~Rodriguez}
\author[lbnl]{G.~Ron}
\author[tjnaf]{A.~Saha} 
\author[smu]{A.~J.~Sarty}
\author[uva,tmu]{B.~Sawatzky}
\author[rut]{E.~Schulte}
\author[tau]{R.~Shneor}
\author[mit]{N.~Sparveris}
\author[ksu]{R.~Subedi}
\author[usc]{S.~Strauch}
\author[tjnaf]{V.~Sulkosky}
\author[uil]{Y.~Wang}
\author[tjnaf]{B.~Wojtsekhowski}
\author[snu]{X.~Yan}
\author[tmu]{H.~Yao}
\author[mit]{X.~Zhan}
\author[uva]{X.~Zheng}

\address[tau]{Tel Aviv University, Tel~Aviv 69978, Israel} 
\address[ukn]{University of Kentucky, Lexington, Kentucky 40506, USA}
\address[nrcn]{NRCN, P.O.Box 9001, Beer-Sheva 84190, Israel}
\address[gwu]{George Washington University, Washington D.C. 20052, USA}
\address[fiu]{Florida International University, Miami, Florida 33199, USA}
\address[tjnaf]{Thomas Jefferson National Accelerator Facility, Newport News, Virginia 23606, USA}
\address[odu]{Old Dominion University, Norfolk, Virginia 23508, USA}
\address[uva]{University of Virginia, Charlottesville, Virginia 22904, USA}
\address[infn]{INFN, gruppo collegato Sanit\`a and Istituto Superiore di Sanit\`a, Department TESA, I-00161 Rome, Italy}
\address[rut]{Rutgers, The State University of New Jersey, Piscataway, New Jersey 08855, USA}
\address[smu]{Saint Mary's University, Halifax, Nova Scotia B3H 3C3, Canada}
\address[dal]{Dalhousie University, Halifax, Nova Scotia B3H 3J5, Canada}
\address[ksu]{Kent State University, Kent, Ohio 44242, USA}
\address[snu]{Seoul National University, Seoul 151-747, Korea}
\address[cwm]{College of William and Mary, Williamsburg, Virginia 23187, USA}          
\address[duu]{Duke University, Durham, NC 27708, USA}
\address[mit]{Massachusetts Institute of Technology, Cambridge, Massachusetts 02139, USA}
\address[lbnl]{Lawrence Berkeley National Laboratory, Berkeley, California 94720, USA}
\address[tmu]{Temple University, Philadelphia, Pennsylvania 19122, USA}
\address[usc]{University of South Carolina, Columbia, South Carolina 29208, USA}
\address[uil]{University of Illinois at Urbana-Champaign, Urbana, IL 61801, USA}

%\collaboration{The Jefferson Lab Hall A Collaboration}
%\noaddress 
\begin{abstract}
We present a study of high energy photodisintegration of proton-pairs through the $\gamma+{\rm ^3He} \to p+p+n$
channel.  Photon energies, $E_\gamma$, from 0.8 to 4.7~GeV were used in kinematics corresponding to 
a proton pair with high relative momentum and a neutron nearly at rest.
The $s^{-11}$ scaling of the cross section, as predicted by
the constituent counting rule for two nucleon photodisintegration, 
was observed for the first time.  
The onset of the scaling is at a higher
energy and the cross section is significantly lower than for deuteron ($pn$ pair)
photodisintegration.  For $E_\gamma$ below the 
scaling region, the scaled cross section was found to present a strong energy-dependent structure not 
observed in deuteron photodisintegration.
\end{abstract}

\end{frontmatter}

%%%%%%%%%%%%%%%%%% introduction %%%%%%%%%%%%%%%%%%%%%%%%

 A common problem in describing  quantum mechanical systems is identifying 
the relevant degrees of freedom needed to efficiently describe the 
underlying reaction dynamics. Conventional nuclear physics descriptions
use meson-baryon degrees of freedom, and it is an ongoing  challenge of modern 
nuclear physics to identify phenomena in which the underlying
quark-gluon degrees of freedom are  important for their description.
In exclusive nuclear reactions, no apparent phase transitions have been
identified which make clear that the relevant degrees of freedom have changed
from hadrons to quarks and gluons.
Hard two-body processes, where all Mandelstam variables $s$, $-u$, and $-t$  
are larger than the $\Lambda_{QCD}^2$ scale are natural 
candidates to reflect the quark substructure of the hadrons and nuclei, since
they involve short distance scales.

Extensive studies of high-energy deuteron photodisintegration
over the past two decades have probed the limits of meson-baryon
descriptions of nuclei and 
reactions~\cite{Napolitano:1988uu,Freedman:1993nt,Belz:1995ge,Bochna:1998ca,Schulte:2001se,Schulte:2002tx,Mirazita:2004rb,Rossi:2004qm},
and the effects of the underlying quark-gluon degrees of freedom. At low 
energies, up through the region of $\Delta$ resonance excitation, 
photodisintegration of the deuteron is well understood, although certain detailed 
problems  remain~\cite{Arenhoevel:2002kh,Schwamb:2001jy,Schwamb:1999qd,Schiavilla:2005zt}.
The calculations are based on meson-baryon degrees of freedom,
constrained by data on $NN$ scattering and pion photo-production~\cite{Arenhoevel:2002kh,Schwamb:2001jy,Schwamb:1999qd}.

Above $\sim$1 GeV, deuteron photodisintegration at large angles leads to 
a large total cm energy and large transverse momenta - this is the hard 
photodisintegration regime. At these high energies photodisintegration
cross sections have been shown to follow the constituent counting rules~\cite{Rossi:2004qm,PhysRevLett.37.269,PhysRevD.14.3003,Brodsky:1973kr,Matveev:1973ra}, that have been been
re-derived from quantum chromodynamics (QCD) and string theory, using the 
Anti-de Sitter / Conformal Field Theory
(AdS/CFT) correspondence~\cite{Brodsky:1973kr,Lepage:1980fj,Polchinski:2001tt}.
Since the $s$-dependence of the cross section at fixed c.m.\ scattering angle\footnote{
Generally, $d\sigma/dt \sim s^{-n}$ where the exponent $n$ is two less than the 
number of point-like constituents in the initial and final states. 
For $\gamma N N \to N N$, $n = 13 - 2 = 11$.}
would naturally arise from the
underlying quark degrees of freedom, the behavior suggests that the quarks are 
the relevant degrees of freedom.
Furthermore,  meson-baryon calculations cannot handle the hundreds of available
resonance channels that can be excited, and quark degrees of freedom naturally
sum over the baryon resonances~\cite{Gilman:2001yh}. Several quark model 
calculations have been used to explain the behavior of high-energy
photodisintegration~\cite{Brodsky:1983kb,Grishina:2001cr,Frankfurt:1999ik},
and moderate success has even been achieved in explaining some polarization
observables~\cite{Wijesooriya:2001yu,Jiang:2007ge,Sargsian:2003sz,Grishina:2001cr,Frankfurt:1999ik}.

In an attempt to more clearly
identify the underlying dynamics at play, we present in this work the
first high-energy measurement of photodisintegration of two protons, using ${\rm ^3He}$.
 In these  measurements the transverse momentum
of the protons exceeded 1 GeV/$c$. The basic idea is that
theoretical models should be able to predict the relative size of $pp$
versus $pn$ disintegration~\cite{Brodsky:2003ip}. Also, if the $pp$ and $pn$ disintegration are
related to the corresponding $pp$ and $pn$ elastic scattering via hard re-scattering, the differences
in the elastic scattering should be reflected in corresponding
differences in the photodisintegration processes.
Finally, the relative smallness of the low-energy $\gamma$-$pp$
disintegration process, compared to $\gamma$-$pn$, has been explained as 
resulting from the small magnetic moment of the $pp$ pair~\cite{Laget:1988cq}.
One of the motivations for this measurement is to check if this behavior continues at higher energies.

%%%%%%%%%%%%%%%%%% Experimental setup %%%%%%%%%%%%%%%%%%%%%%%%

\begin{figure}[htb]
\includegraphics[width=\linewidth]{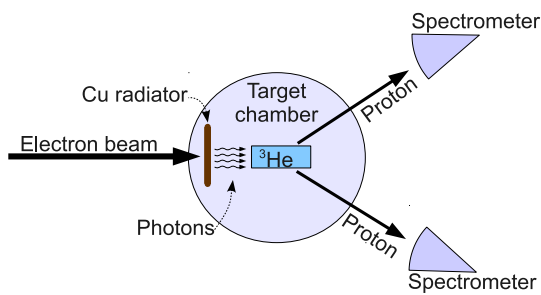}
\caption{\label{fig:exp_setup} (Color online) Experimental setup: Bremsstrahlung photons 
generated in a copper radiator by an electron beam impinged on a ${\rm ^3He}$ gas target. 
Protons were detected with the spectrometers. Elements are not to scale.}
\end{figure}

The experiment (E03-101) ran in Hall A of the Thomas Jefferson National Accelerator Facility (JLab)~\cite{Alcorn:2004sb}. 
The experimental setup is schematically described in Fig.~\ref{fig:exp_setup}.
Bremsstrahlung photons were generated when the electron beam with energy 0.8, 1.1, 1.7, 2.1, 2.5, 3.1, 4.1 or 4.7 GeV
impinged on a copper radiator. The 6\%-radiation-length radiator was located in the scattering 
chamber 38~cm upstream of the center of a 20~cm long cylindrical ${\rm ^3He}$ gas target, with density of 0.079\ g/cm$^3$.
The size of the photon beam spot on the target, $\sim 2$ mm, is dominated by electron
beam rastering intended to distribute the heat load across the targets,
and is much smaller than the $\sim 1$ cm size target windows and apertures.  
Protons from the target were detected with the Hall A high-resolution spectrometers (HRSs) 
at kinematic settings corresponding to the $\theta_{c.m.}$ = 90$^{\circ}$ photodisintegration 
of a proton pair at rest. For each spectrometer, the scattering angles, momenta, 
and interaction position at the target were reconstructed from trajectories measured 
with Vertical Drift Chambers (VDCs) located in the focal
plane.  Two planes of plastic scintillators provided triggering and time-of-flight information for particle identification.

\begin{figure}[htb]
\includegraphics[width=\linewidth]{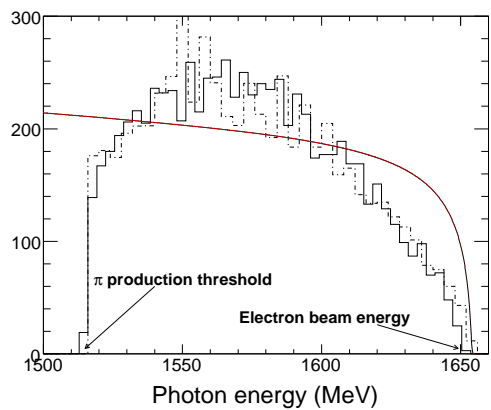}
\caption{\label{fig:egamma_reconst} (Color online) Photon energy distributions for an electron beam of 1.655 GeV. 
Solid line represent the reconstructed photon energy of the photodisintegration events, dashed line the 
simulation results. The calculated bremsstrahlung spectrum~\cite{Matthews:1973xn} is shown as a solid red line. 
The difference between the data / Monte Carlo and the bremsstrahlung shapes
results from the spectrometer acceptances and resolutions plus the energy dependence of the cross section.
All distributions are normalized to the measured yield.}
\end{figure}

The incident photon energy, as well as the neutron's recoil momentum were reconstructed from the momentum 
and angles of the scattered protons under the assumption of $ppn$ final-state kinematics. 
In order to assure the validity of this assumption,  only events between the bremsstrahlung 
endpoint and the pion production threshold were used in the analysis. Figure~\ref{fig:egamma_reconst} 
shows the photon energy distributions for an electron beam total energy of 1655~MeV.

%%%%%%%%%%%%%%%%%% Data analysis %%%%%%%%%%%%%%%%%%%%%%%%

Proton pairs produced in coincidence can result from either a photon or an electron 
disintegrating the ${\rm ^3He}$ nucleus. 
We took data with the radiator in and out of the beam, to extract the number of events resulting 
from photons produced in the bremsstrahlung radiator. As in~\cite{Schulte:2001se}, 
due to low rates, measurements 
without the radiator were taken only up to $E_\gamma$ = 3.1 GeV. As theoretical
guidance~\cite{Wright:1982vb,Tiator:1982if} indicates that the ratio of electro- to photo-disintegration should vary slowly with energy, the correction for 
higher photon energies was extrapolated from the measurements at the lower energies.
With the momentum and path well determined by the narrow spectrometer acceptances, 
protons were selected by cutting on the time of flight.  The reconstructed reaction point 
location was selected to be within the central 10 cm of the target. Random events 
were removed with  narrow cuts on the coincidence time and the difference between 
the reaction points, independently reconstructed for each spectrometer.  A photon energy 
cut was placed as described above. A cut was placed on the reconstructed neutron momentum 
to be less than 100~MeV/$c$. With these measurement conditions, the neutron in the $^3$He 
can be considered, at least approximately, as a static spectator~\cite{Brodsky:2003ip,Niccolai:2004ne}. 
Correction for the finite acceptance of the spectrometers 
was done using the standard Hall A Monte-Carlo simulation software MCEEP~\cite{MCEEP}. 

%\begin{figure}[!t]
%  \begin{minipage}[t]{1.\linewidth}
%\begin{center}
%\includegraphics[width=3.25in]{dsdt_noclas.eps}
%\caption[]{\label{fig:results} (Color online) The d($\gamma$,$p$)$n$ (a) and $^3$He($\gamma$,$pp$)$n$ (b) invariant cross section, $^3$He($\gamma$,$pp$)$n$ events were selected with $p_n<100$ MeV/$c$.  d($\gamma$,$p$)$n$ results are taken from \cite{gammad_mainz,Napolitano:1988uu,Freedman:1993nt,Belz:1995ge,Bochna:1998ca}. Model predictions are taken from \cite{Brodsky:2003ip}, \cite{Sargsian:2008zm}. Error bars represent statistical uncertainty. The fits and their parameters are shown in solid lines.}
%  \end{minipage}
%\end{center}
%\end{figure}

%%%%%%%%%%%%%%%%%% Results %%%%%%%%%%%%%%%%%%%%%%%%

\begin{figure}[!t]
\includegraphics[width=3.25in]{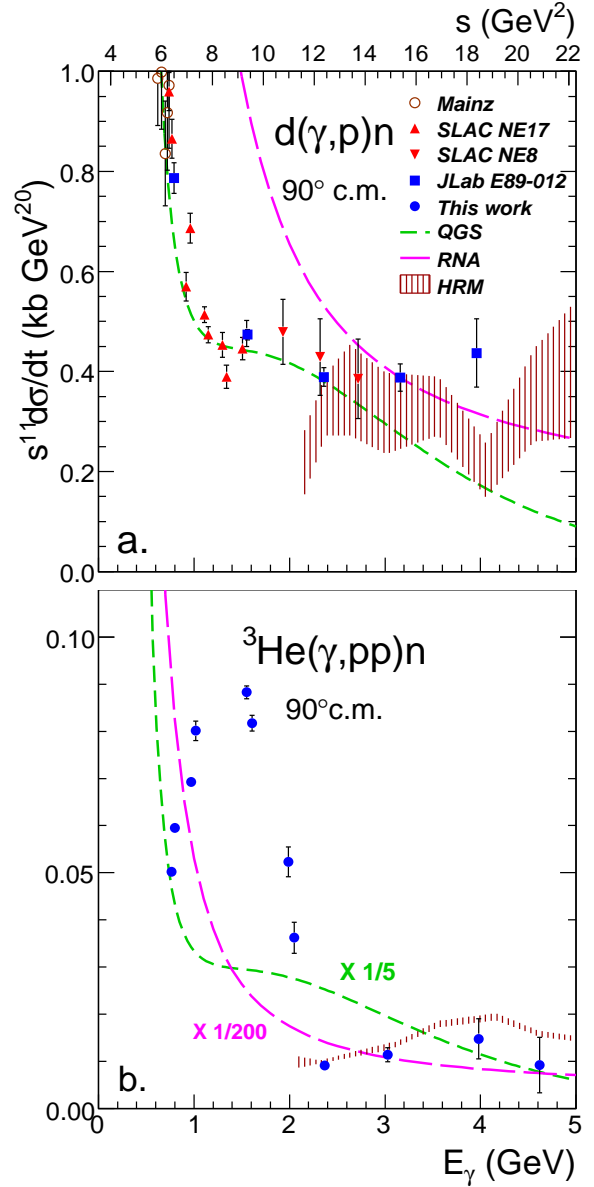}
\caption[]{\label{fig:results_scaled} (Color online) 
Invariant cross section scaled by $s^{11}$ for d($\gamma$,$p$)$n$ (a), taken from previous 
work~\cite{Napolitano:1988uu,Freedman:1993nt,Belz:1995ge,Bochna:1998ca,Schulte:2001se,
Schulte:2002tx,Mirazita:2004rb,Rossi:2004qm} and for $^3$He($\gamma$,$pp$)$n$ (b), from this work.  
The $^3$He($\gamma$,$pp$)$n$ events were selected with $p_n<100$ MeV/$c$. Up to 2.1 GeV, the photon energy bins are 70 MeV, and above it 140 MeV. Model predictions are taken from~\cite{Brodsky:2003ip,Sargsian:2008zm}. In (b), RNA is divided by a factor of 200 and QGS by a factor of 5 to be shown on this scale. Only the statistical uncertainty
is shown.}
\end{figure}

Figure~\ref{fig:results_scaled} shows the $\gamma+{\rm d} \to p+n$ and $\gamma+{\rm ^3He} \to p+p+n$ $\theta_{c.m.}$ = 90$^{\circ}$ cross section scaled with $s^{11}$. The $^3$He($\gamma$,$pp$)$n$ events were selected with $p_n<100$~MeV/$c$. 
A fit of the $\theta_{c.m.}$ = 90$^{\circ}$ cross section to $A\times s^n$ yielded $n = -11.1 \pm 0.1$ for the deuteron for $E_\gamma\ \sim$~1 GeV. For the $pp$ pairs n=$-10.5 \pm 0.6$ and the scaling commences at $E_\gamma\ \approx$ 2.2 GeV. In the scaling region, the $\theta_{c.m.}$ = 90$^{\circ}$ cross section for $^3$He($\gamma$,$pp$)$n$ is about 40~times smaller than that 
for $d$($\gamma$,$p$)$n$.
Correcting for the abundance of $pp$ pairs with low neutron momentum ($p_n<100$ MeV/$c$) as calculated from~\cite{Schiavilla:2005zt}, 
yielded $\sim$ 20 times smaller cross section for disintegration of a $pp$ pair in $^3$He than of a $pn$ pair in the deuteron. 
The cross section is compared to theoretical models, discussed below, that produced predictions for the photodisintegration of the deuteron   
and a $pp$ pair in $^3$He~\cite{Brodsky:2003ip}. 

The statistical uncertainty rises with the incident photon energy from less than 1\% to 64\%.
The systematic error for all data is between 5 and 10\%. At low energy, the systematic error
dominates with the major contribution due to 
the large number of $\theta_{c.m.}$ = 90$^{\circ}$ proton pairs not detected by the spectrometers. 
This acceptance limitation is handled by the simulation, but introduces a larger systematic uncertainty.
At high energy the systematic uncertainty is dominated by the $^3$He electro-disintegration subtraction (due to the 
extrapolation from lower energies).

The photodisintegration of ${\rm ^3He}$ has also been measured with the Hall B/CLAS~\cite{Mecking:2003zu} detector at Jefferson Lab,
using tagged photons of 0.35 to 1.55 GeV~\cite{Niccolai:2004ne}. The large acceptance of the spectrometer 
allowed detection of the two outgoing protons over a wide range of momentum and angles. Events corresponding 
to $\theta_{c.m.}$ = 90$^{\circ}$ break-up of proton pairs were selected with various cuts on neutron momentum.  
Preliminary yet unpublished single differential cross-section results from CLAS~\cite{priv:steffen} in the range of 0.85 GeV $<\ E_\gamma\ <$ 1.1 GeV agree within 10\% 
with the data presented here.

%%%%%%%%%%%%%%%%%% Discussion %%%%%%%%%%%%%%%%%%%%%%%%

Our new data along with previous low-energy data indicate that the $^3$He two-proton disintegration can be divided into three energy regions. At \textit{low} photon energies (below $E_\gamma\ \approx$ 0.5 GeV), the dynamics of $\theta_{c.m.}$ = 90$^{\circ}$ proton-pair 
breakup is governed by hadron and meson degrees of freedom and the cross section has a large three-body component~\cite{Laget:1984yy}.

In a \textit{transition} region (1 GeV $<\ E_\gamma\ <$ 2.2 GeV) the scaled cross section for deuteron ($pn$ pairs) breakup is flat while for $pp$ 
pairs a significant structure is observed. This structure may be the result of resonances in the $\gamma$N or $\gamma$NN systems. 
The energy dependence in the transition region more closely
resembles the energy behavior of the photo-induced
pion production~\cite{Besch:1982sx,Benz:1974tt,Zhu:2004dy} than that of deuteron photo-disintegration. It has been suggested that the structure might result from a meson photo-produced on a proton and then absorbed on a $pn$ pair~\cite{priv:laget}.

In the \textit{scaling} region the cross section for both deuteron ($pn$) and $pp$ breakup scales in agreement with the constituent counting 
rule~\cite{Brodsky:1973kr,Lepage:1980fj,Polchinski:2001tt}. 
For proton-pair break-up, the onset of the scaling is at $E_\gamma\ \approx$ 2.2 GeV, while for deuteron ($pn$ pair) scaling commences at $E_\gamma$ $\approx$~1 GeV~\cite{Schulte:2002tx}. The scaling in the $^3$He case indicates that in this regime the two-body process is dominant. It further suggests (in a relatively model-independent way)
that the relevant degrees of freedom that govern the dynamics are the quarks. In a hadronic picture, two-body/one-step
processes are strongly suppressed since no charge can be exchanged between the protons.

The reduced nuclear amplitude (RNA) formalism~\cite{Brodsky:1983kb}  after
normalization to the deuteron data~\cite{Brodsky:2003ip} yields cross sections that are about 200~times
larger than the present data. The quark-gluon string model (QGS)~\cite{Grishina:2001cr,Grishina:2002ph}, 
as estimated in~\cite{Brodsky:2003ip}, predicts 
cross sections about a factor of 5 larger than measured. The QCD hard 
re-scattering model (HRM)~\cite{Frankfurt:1999ik} allows an absolute calculation of the 
cross sections for both $pn$ and $pp$ pair photodisintegration 
from nucleon-nucleon measured cross sections without adjustable parameters. 
It reproduces reasonably well the deuteron data and the proton pair cross section.

An explanation for the low magnitude of the scaled cross section of proton-pair breakup is given in the HRM~\cite{Sargsian:2008zm} 
by a cancellation of the opposite sign of the NN helicity amplitudes $\phi_3$ and $\phi_4$
in the $pp$ breakup~\footnote{$\phi_3$ and $\phi_4$ are the NN elastic scattering helicity amplitudes that connect zero 
helicity in the initial states to zero helicity in the final state.$\phi_3$ does it with no helicity 
exchange. $\phi_4$ exchanges helicity between the scattered nucleons. 
This cancellation of $\phi_3$ and $\phi_4$ was not recognized in~\cite{Brodsky:2003ip}.}. 
The energy dependence predicted by the HRM in the scaling region agrees well with the data. Therefore,
hard re-scattering is a plausible explanation for the origin of the large transverse momenta.
Models that hold compact NN pairs in the initial state to be the reason for the 
large transverse momenta~\cite{Brodsky:1983kb} would have to assume either a fairly 
low abundance of $pp$ pairs within the ${\rm ^3He}$ wave function or the same 
type of nuclear amplitude cancellation in order to explain the low magnitude 
of the $pp$ break-up scaled cross section. 

Another possible explanation for the cross-section magnitude may lie in 
tensor correlations~\cite{Sargsian:2005ru,Schiavilla:2006xx,Alvioli:2007zz}.
These nucleon-nucleon correlations cause the ratio of $pp$ to $np$ pairs to be $\sim$5\% in the relative momentum 
range of 300-600 MeV/$c$ for both high-energy electron and proton scattering~\cite{Piasetzky:2006ai,Shneor:2007tu,Subedi:2008zz}. 
Starting with such a pair and final state re-scattering might lead to the observed 
relative transverse momentum and would explain the relatively small cross sections.

In conclusion, we have presented the first high-energy measurements of
$pp$ photodisintegration through the $\gamma+{\rm ^3He} \to p+p+n$ reaction.
For energies between about 1 and 2~GeV,
the cross section shows a large structure, possibly related
to excitation of baryon resonances.
Above about 2 GeV, the measured cross section scales as $s^{-11}$, but at a
level about 20~times smaller than the deuteron disintegration cross section.
This arises naturally from the hard rescattering model due to cancellation
of $pp$ scattering amplitudes. Other models tend to over-predict the
$pp$ disintegration cross section.
If the underlying dynamics of photodisintegration are sensitive to nucleon
pairs in the relative momentum range 300 - 600 MeV/c, then an alternative
explanation for the relative cross-section magnitude of  $\gamma$d to $\gamma pp$ arises
from tensor correlations.

%\begin{acknowledgements}

We thank M. M. Sargsian for initiating and escorting this study 
and S. J. Brodsky, L. L. Frankfurt and M. Strikman for helpful discussions.
We thank the JLab physics and accelerator divisions for their support and especially the 
CLAS collaboration of Hall B, for allowing us access to their data. 
This work was supported by the U.S.\ Department of Energy,
the U.S.\ National Science Foundation, the Israel Science Foundation, 
and the US-Israeli Bi-National Scientific
Foundation. Jefferson Science Associates operates
the Thomas Jefferson National Accelerator Facility under DOE
contract DE-AC05-06OR23177.
%\end{acknowledgements}

\bibliographystyle{elsarticle-num}
\bibliography{gammaPP-v20}

\begin{thebibliography}{10}
\expandafter\ifx\csname url\endcsname\relax
  \def\url#1{\texttt{#1}}\fi
\expandafter\ifx\csname urlprefix\endcsname\relax\def\urlprefix{URL }\fi
\expandafter\ifx\csname href\endcsname\relax
  \def\href#1#2{#2} \def\path#1{#1}\fi

\bibitem{Napolitano:1988uu}
J.~Napolitano, et~al., Phys. Rev. Lett. 61 (1988) 2530.

\bibitem{Freedman:1993nt}
S.~J. Freedman, et~al., Phys. Rev. C 48 (1993) 1864.

\bibitem{Belz:1995ge}
J.~E. Belz, et~al., Phys. Rev. Lett. 74 (1995) 646.

\bibitem{Bochna:1998ca}
C.~Bochna, et~al., Phys. Rev. Lett. 81 (1998) 4576.

\bibitem{Schulte:2001se}
E.~C. Schulte, et~al., Phys. Rev. Lett. 87 (2001) 102302.

\bibitem{Schulte:2002tx}
E.~C. Schulte, et~al., Phys. Rev. C 66 (2002) 042201.

\bibitem{Mirazita:2004rb}
M.~Mirazita, et~al., Phys. Rev. C 70 (2004) 014005.

\bibitem{Rossi:2004qm}
P.~Rossi, et~al., Phys. Rev. Lett. 94 (2005) 012301.

\bibitem{Arenhoevel:2002kh}
H.~Arenhoevel, E.~M. Darwish, A.~Fix, M.~Schwamb, Mod. Phys. Lett. A18 (2003)
  190.

\bibitem{Schwamb:2001jy}
M.~Schwamb, H.~Arenhovel, Nucl. Phys. A696 (2001) 556.

\bibitem{Schwamb:1999qd}
M.~Schwamb, H.~Arenhovel, Nucl. Phys. A690 (2001) 647.

\bibitem{Schiavilla:2005zt}
R.~Schiavilla, Phys. Rev. C 72 (2005) 034001.

\bibitem{PhysRevLett.37.269}
S.~J. Brodsky, B.~T. Chertok, Phys. Rev. Lett. 37 (1976) 269.

\bibitem{PhysRevD.14.3003}
S.~J. Brodsky, B.~T. Chertok, Phys. Rev. D 14 (1976) 3003.

\bibitem{Brodsky:1973kr}
S.~J. Brodsky, G.~R. Farrar, Phys. Rev. Lett. 31 (1973) 1153.

\bibitem{Matveev:1973ra}
V.~A. Matveev, R.~M. Muradian, A.~N. Tavkhelidze, Nuovo Cim. Lett. 7 (1973)
  719.

\bibitem{Lepage:1980fj}
G.~P. Lepage, S.~J. Brodsky, Phys. Rev. D 22 (1980) 2157.

\bibitem{Polchinski:2001tt}
J.~Polchinski, M.~J. Strassler, Phys. Rev. Lett. 88 (2002) 031601.

\bibitem{Gilman:2001yh}
R.~Gilman, F.~Gross, J. Phys. G28 (2002) R37.

\bibitem{Brodsky:1983kb}
S.~J. Brodsky, J.~R. Hiller, Phys. Rev. C 28 (1983) 475.

\bibitem{Grishina:2001cr}
V.~Y. Grishina, et~al., Eur. Phys. J. A10 (2001) 355.

\bibitem{Frankfurt:1999ik}
L.~L. Frankfurt, G.~A. Miller, M.~M. Sargsian, M.~I. Strikman, Phys. Rev. Lett.
  84 (2000) 3045.

\bibitem{Wijesooriya:2001yu}
K.~Wijesooriya, et~al., Phys. Rev. Lett. 86 (2001) 2975.

\bibitem{Jiang:2007ge}
X.~Jiang, et~al., Phys. Rev. Lett. 98 (2007) 182302.

\bibitem{Sargsian:2003sz}
M.~M. Sargsian, Phys. Lett. B587 (2004) 41.

\bibitem{Brodsky:2003ip}
S.~J. Brodsky, et~al., Phys. Lett. B578 (2004) 69.

\bibitem{Laget:1988cq}
J.~M. Laget, Nucl. Phys. A497 (1989) 391c.

\bibitem{Alcorn:2004sb}
J.~Alcorn, et~al., Nucl. Instrum. Meth. A522 (2004) 294.

\bibitem{Matthews:1973xn}
J.~L. Matthews, R.~O. Owens, Nucl. Instrum. Meth. 111 (1973) 157.

\bibitem{Wright:1982vb}
L.~e. Wright, L.~Tiator, Phys. Rev. C 26 (1982) 2349.

\bibitem{Tiator:1982if}
L.~Tiator, L.~E. Wright, Nucl. Phys. A379 (1982) 407.

\bibitem{Niccolai:2004ne}
S.~Niccolai, et~al., Phys. Rev. C 70 (2004) 064003.

\bibitem{MCEEP}
P.~E. Ulmer{ CEBAF-TN-91-101}.

\bibitem{Sargsian:2008zm}
M.~M. Sargsian, C.~Granados, Phys. Rev. C 80 (2009) 014612.

\bibitem{Mecking:2003zu}
B.~A. Mecking, et~al., Nucl. Instrum. Meth. A503 (2003) 513.

\bibitem{priv:steffen}
S.~Strauch, private communication.

\bibitem{Laget:1984yy}
J.~M. Laget, Phys. Lett. B151 (1985) 325.

\bibitem{Besch:1982sx}
H.~J. Besch, F.~Krautschneider, K.~P. Sternemann, W.~Vollrath, Z. Phys. C16
  (1982) 1.

\bibitem{Benz:1974tt}
P.~Benz, et~al., Nucl. Phys. B65 (1973) 158.

\bibitem{Zhu:2004dy}
L.~Y. Zhu, et~al., Phys. Rev. C 71 (2005) 044603.

\bibitem{priv:laget}
J.~M. Laget, private communication.

\bibitem{Grishina:2002ph}
V.~Y. Grishina, et~al., Eur. Phys. J. A18 (2003) 207.

\bibitem{Sargsian:2005ru}
M.~M. Sargsian, T.~V. Abrahamyan, M.~I. Strikman, L.~L. Frankfurt, Phys. Rev. C
  71 (2005) 044615.

\bibitem{Schiavilla:2006xx}
R.~Schiavilla, R.~B. Wiringa, S.~C. Pieper, J.~Carlson, Phys. Rev. Lett. 98
  (2007) 132501.

\bibitem{Alvioli:2007zz}
M.~Alvioli, C.~Ciofi~degli Atti, H.~Morita, Phys. Rev. Lett. 100 (2008) 162503.

\bibitem{Piasetzky:2006ai}
E.~Piasetzky, M.~Sargsian, L.~Frankfurt, M.~Strikman, J.~W. Watson, Phys. Rev.
  Lett. 97 (2006) 162504.

\bibitem{Shneor:2007tu}
R.~Shneor, et~al., Phys. Rev. Lett. 99 (2007) 072501.

\bibitem{Subedi:2008zz}
R.~Subedi, et~al., Science 320 (2008) 1476.

\end{thebibliography}

\end{document}